\documentstyle[12pt,epsfig]{article}
\newcommand{\bm}[1]{\mbox{\boldmath $#1$}}
\def\be{\begin{equation}}
\def\ee{\end{equation}}

\title{Different transport regimes in a spatially-extended
 recirculating background}

\author{P.~Castiglione$^{1,2}$, R.~Festa$^{3}$ and  A.~Mazzino$^{3,4}$\\
\small{$^{1}$
Laboratoire de Physique Statistique, Ecole Normale Sup\'erieure, Paris
Cedex 05, France .}\\
\small{$^{2}$ INFM,
Dipartimento di Fisica, Universit\`a ``La Sapienza'', I--00185
Roma, Italy.}\\
\small{$^{3}$ INFM,
Dipartimento di Fisica, Universit\`a di Genova, I--16146
Genova, Italy.}\\
\small$^4$
The Niels Bohr Institute, Blegdamsvej 17, DK-2100 Copenhagen, Denmark .}

\begin{document}
\maketitle
\date{}
\vspace*{-0.6cm}
\begin{abstract}
Passive scalar transport in a spatially-extended background
of roll convection is considered in the time-periodic regime.
The latter arises due to the even oscillatory instability of the
cell lateral boundary, here accounted for by
sinusoidal oscillations of
frequency $\omega$. By varying the latter parameter, the strength of
anticorrelated regions of the velocity field can be controled and
the conditions under which either an enhancement or a reduction of transport
takes place can be created. Such two ubiquitous regimes are triggered by a
small-scale(random) velocity field superimposed to the recirculating
background.
The crucial point is played by the dependence of Lagrangian
trajectories on the statistical properties of the small-scale
velocity field, e.g.~its correlation time or its energy.
\end{abstract}

\noindent PACS 47.27Qb -- Turbulent diffusion.

\vspace{4mm}

Transport in turbulent flows with recirculation is a problem of great
interest both in the atmosphere and in the ocean
\cite{RY82,P88,RJR95}.
In the atmosphere, the so-called horizontal roll vortices or, briefly, rolls
are the paradigm of recirculating pattern.
They can be considered as a manifestation of B\'enard--Rayleigh convection
occurring when conditions of combined surface
heating and strong winds take place in the atmospheric boundary layer. 
Their depth
equals the mixing layer and the ratio of lateral to vertical dimensions for
a roll pair is about 3:1 (see, e.g., Ref.~\cite{B70}).
In the following, such type of flow configuration will be the background
where particle tracers (i.e.~a passive scalar) will be plunged and their
statistics investigated.\\
An important characteristic of atmospheric rolls is their
large spatial extension which makes possible the description of the related
dispersion problem through an effective diffusion equation, i.e.~a Fick
equation
for the large-scale slow-varying passive scalar concentration
where the molecular diffusivity is replaced by an enhanced
(eddy-) diffusivity.\\
The existence of an asymptotic diffusive regime for the large scale
concentration can be rigorously proved by using, e.g., multiscale techniques
\cite{BLP78}.  A simple  Lagrangian interpretation of the asymptotic
diffusive regimes is based on ``central limit'' argumentations.
For Lagrangian chaotic flows, velocity correlations decay fastly and,
as a consequence, the particle displacement, $\delta x(t)$, at the time $t$,
is the results
of the sum of almost independent advecting contributions. The result
is that $\delta x(t)$ undergoes Brownian
motion when observed on times larger than the typical velocity
correlation time.

In this asymptotic framework,
the effect on the dispersion process of the small-scale components of the
velocity field is the renormalization of the effective diffusion coefficient.
Notice that, as pointed out in Ref.~\cite{CCVZ99}, an
eddy-diffusivity based description should not be possible in the presence
of finite-size domains with a small number of recirculations
where, necessarily, the asymptotic regime might not be reached and the
dynamics is governed by transient behaviors \cite{Y80}. Interesting studies
on this regimes can be found, e.g., in Refs.~\cite{SD96,ABCCV97,KT97}.

For the system we are going to investigate, the
characteristic length of the organized array of cells is
smaller than the size of the domain.
As a consequence, the eddy-diffusivity tensor,
defined through the following asymptotic limit,
\begin{equation}
D_{\alpha\beta}^E = lim_{t\to\infty}\frac{1}{2\;t}\langle
[x_{\alpha}(t)-x_{\alpha}(0)] [x_{\beta}(t)-x_{\beta}(0)]\rangle,
\label{diffu}
\end{equation}
with $x(t)$ being the particle position at the time $t$, and
$\langle\cdot\rangle$ the average over an ensemble of tracer particles,
turns out to be a well-defined mathematical quantity.

Atmospheric rolls show a wide range of regimes ranging from (almost)
time independent to turbulent flow. The related dispersion
phenomena are clearly strongly influenced by these different regimes and,
as a consequence, transport rates vary over a wide range. An intermediate
regime attracting considerable attention both theoretically \cite{CNRY91},
experimentally \cite{SG88} and numerically \cite{CMMV99} is the time-periodic
regime, where the transport process is dominated by advection
of tracer particles across the lateral boundary. Such regime will be the
main concern of the present Letter.
Specifically, the main question addressed here concerns the role of the
small-scale
(not explicitly resolved) components of the velocity field on the large-scale
transport.
We shall show that the superposition of a colored (random) noise velocity
(that can be though as associated to a small-scale turbulent motion with
nonvanishing memory) to the convective (deterministic) background strongly
affects the transport process: either an enhanced or a reduced (with
respect to
the white case) eddy-diffusion may occur, depending
on the frequency of the lateral roll oscillations.
The interference mechanism, recently proposed in Refs.~\cite{MV97,MC99}
for the simple, idealized parallel flow, is identified here as the responsible
of this twofold behavior.

Our two-dimensional model for the roll-convection follows Ref.~\cite{SG88}.
Specifically, the convective flow  is defined by the
following stream function\,:
\begin{equation}
\psi (x,y,t) = \psi_0 \sin [k_x(x + B\sin\omega t )] \sin(k_y y) \;\;\; ,
\label{gollu}
\end{equation}
where $y$ ranges from $0$ to $L_y=2\pi/k_y$.
The stream function
(\ref{gollu}) describes single-mode, two-dimensional convection with
rigid boundary condition, where the even oscillatory instability
is accounted for by the term $B\sin \omega t$, representing the
lateral oscillation of the roll.  In Ref.~\cite{SG88}, a quantitative
comparison of the behavior in this flow with the experimental data has
shown that the basic mechanisms of convective transport are well
captured by the expression (\ref{gollu}).

The periodicity of the cell along the $x$-axis
is denoted by $L$  ($L=2\pi/k_x$) while its depth (along the $y-$direction)
is $L/3$ (i.e.~$k_y=3k_x$). The amplitude, $B$, of the roll oscillations
is assumed $\sim 0.13 L$. The
dimensionless parameter controlling the dynamics is
$\epsilon \equiv\omega/\omega_R$, $\omega_R\equiv k_x k_y \psi_0$
being the characteristic frequency of particle oscillations inside the cell.
The two limiting regimes $\epsilon \ll 1$ and $\epsilon
\gg 1$ have been investigated analytically
in Ref.~\cite{CNRY91} to obtain
expressions of the eddy-diffusivity in the limit of zero
molecular diffusivity. Here, we shall concentrate on the behavior for
a wide range of $\epsilon$ and in the presence of
diffusivity. The investigation is however not accessible
by analytical techniques. In order to evaluate eddy-diffusivities,
we have therefore decided to perform
Monte Carlo numerical simulations of the Langevin equation
\begin{equation}
\frac{d\bm{x}(t)}{dt}=\bm{v}(\bm{x}(t)) + \bm{v}'(t)\;\;\;.
\label{lange}
\end{equation}
The velocity field $\bm{v}(\bm{x}(t))$ is incompressible, and related
to the stream-function (\ref{gollu}) through the usual relations
$v_x=\partial_y\psi$, $v_y=-\partial_x\psi$.
The noise term $\bm{v}'(t)$ is a Gaussian, zero-mean
random process with the colored-noise correlation function:
\begin{equation}
\langle v'_\alpha ( t ) \; v'_\beta (t') \rangle =
\frac{D_0}{\tau} \; \delta_{\alpha\beta}
 e^{- \frac{ \mid t-t' \mid }{\tau}}\;\;\; ,
\label{cnoise}
\end{equation}
where $D_0$ can be though as
the (isotropic) eddy-diffusivity arising from the smallest
(not explicitly resolved) scales of turbulent motion and $\tau$ is their
correlation time. Notice that the white-in-time correlation function
is obtained by taking the limit $\tau\to 0$.


 From the Langevin equation (\ref{lange}) the expression (\ref{diffu})
for the eddy diffusivity can be easily rewritten in terms of the Lagrangian
autocorrelation $C_{\alpha \beta}(t)=\langle v_{\alpha}(\bm{x} (t))
v_{\beta}(\bm{x} (0))\rangle$:
\begin{equation}
D_{\alpha\beta}^{E}=D_0 \delta_{\alpha\beta}+\int_0^{\infty}dt \,
C_{\alpha\beta}(t).
\label{Tayl}
\end{equation}

The role played by anticorrelated regions of the velocity field
(i.e.~regions where $C_{\alpha\beta}<0$ in Eq.~(\ref{Tayl}))
on the large-scale transport has been investigated in Refs.~\cite{MV97,MC99}
for the class of parallel flows.
Two
different regimes of transport may occur depending on the extension of such
regions. Specifically, when anticorrelated regions are sufficiently extended
(we denote such regime with the label EAR),
an increasing $\tau$ (for a fixed,
small, $D_0$) causes a reduction of transport (with respect to $\tau=0$)
while an increasing $D_0$ (for a fixed, small, $\tau$) leads to an enhancement
of transport (with respect to $D_0=0$).

The scenario is opposite for anticorrelated regions weak enough (hereafter
regime WAR), that is,  $D_0$ leads to transport reduction while $\tau$ to
transport enhancement. \\
We briefly recall the basic mechanisms characterizing
the above two different regimes.
The first
mechanism works to increase the Lagrangian correlation time
(and thus eddy-diffusivities): this is due to the
fact that $\tau$ makes the particles of diffusing substance
forget their past less rapidly
than in the case $\tau=0$. Thus, the autocorrelation function in
(\ref{Tayl}) decays less rapidly than in the withe-noise case.
This implies an increasing weight of
regions where the velocity is strongly (positively) correlated
and, as an immediate consequence,
an increasing eddy-diffusivity.\\
The second mechanism arises for
flows with closed streamlines and it is
associated to the presence of anticorrelated regions of the velocity field.
The correlation time, $\tau$, is now  working
to increase effects of trapping due to the anticorrelated
zones where the velocity is weak. This means that regions where the velocity
is anticorrelated give an enhanced (again with respect to the withe-noise case)
contribution to the time-integral (\ref{Tayl}).
The contribution of anticorrelated regions to the
time-integral being negative, a reduction of diffusion occurs.\\
\begin{figure}
\begin{center}
\mbox{\psfig{figure=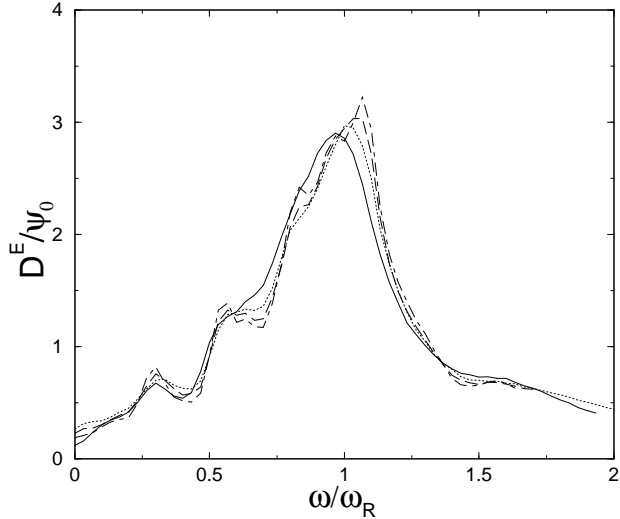,width=.5\linewidth}}
\end{center}
\caption{Behavior of the eddy-diffusivity (component along the direction
of the lateral roll oscillations) vs the lateral roll oscillation frequency.
Full line: $\tau=0$; dotted line: $\tau/t_R=0.24$;
long-dashed line: $\tau/t_R=0.48$;
dot-dashed line: $\tau/t_R=0.95$.}
\end{figure}
Our main aim here is to show that in the presence of roll convection
the aforesaid two mechanisms are relevant and work in competition
thus governing the large-scale transport.
The frequency, $\omega$,
of the lateral roll oscillation is identified here as one of
the parameters controlling the crossover from the WAR and the EAR regimes.
Notice that, unlike Refs.~\cite{MV97,MC99}, our control
parameter, $\omega$, is not trivially related to the extension of
anticorrelated regions. The relation is intrinsic and selected by the dynamics.
Indeed, by varying $\omega$, it is possible to synchronize \cite{CCMVV98} the
frequency (of order of $\omega_R$)
of particles inside the cell
with the frequency, $\omega$, of the lateral roll oscillation.
Due to this synchronization mechanism, the eddy-diffusivity as a function
of $\omega$ can have maxima (when oscillations are in phase) or minima
(when oscillations are in phase opposition).
Moreover from  Eq.~(\ref{Tayl}) it results that maxima (minima)
of diffusion are associated to the
 flow configurations with the weakest (strongest) anticorrelated regions.

In order to evaluate the component of the eddy-diffusivity
along the direction (e.g., the $x$-axis)
of the lateral roll oscillation as a function of $\omega$,
numerical integration of Eq.~(\ref{lange}) has been made by using a
second-order Runge-Kutta scheme and then performing a
linear fit of $\langle [x(t) - x(0)]^2\rangle$ {\it vs } $t$.
Averages are made over different realizations and performed by
uniformly distributing $10^6$ particles in the basic periodic cell.
The system evolution has been
computed up to times $10^4\;t_R$, where $t_R\equiv 2\pi/\omega_R$.

\begin{figure}
\vfill \begin{minipage}{.495\linewidth}
\begin{center}
\mbox{\psfig{figure=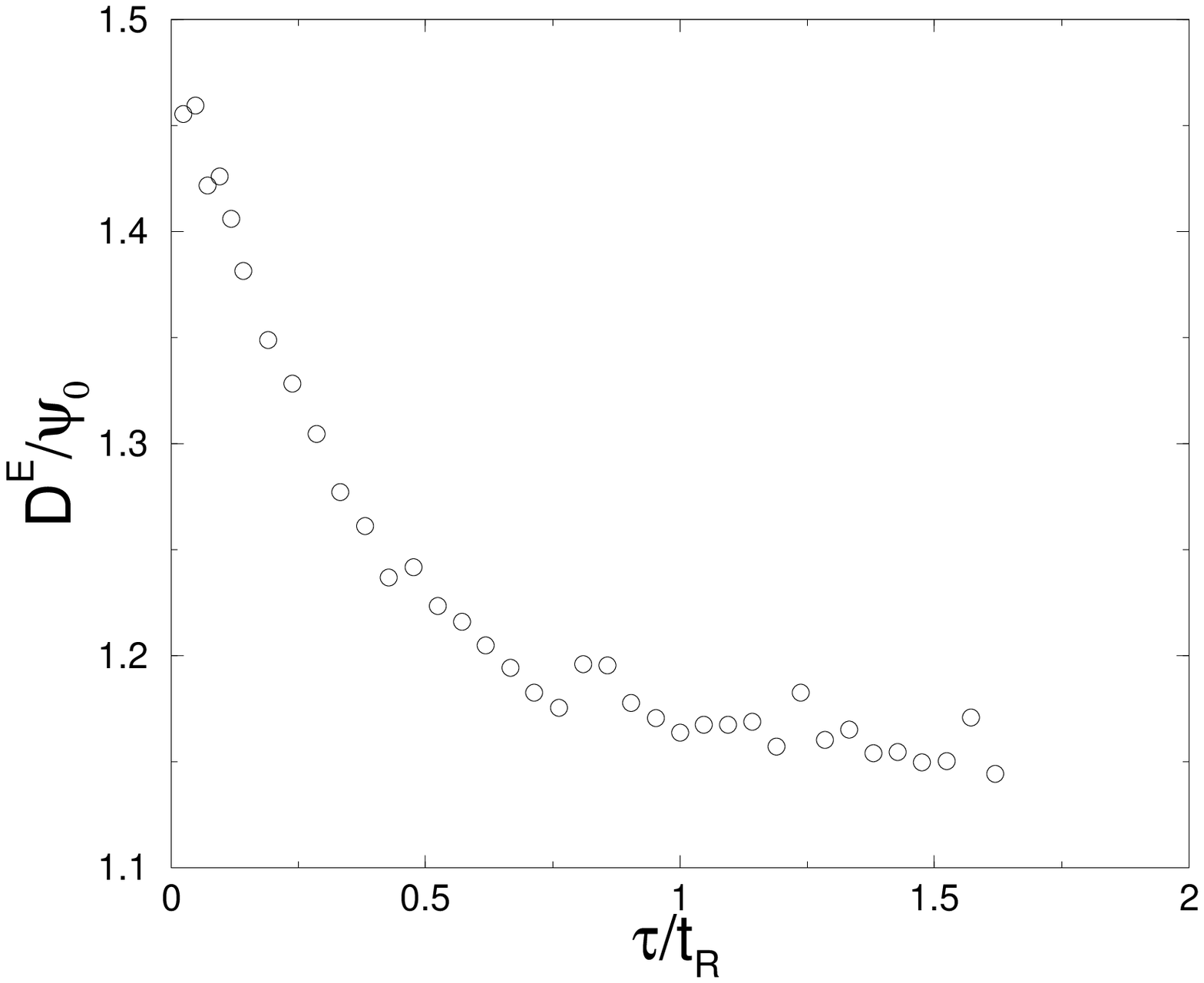,width=.9\linewidth}} 
\end{center}
\end{minipage} \hfill
\begin{minipage}{.495\linewidth}
\begin{center}
\mbox{\psfig{figure=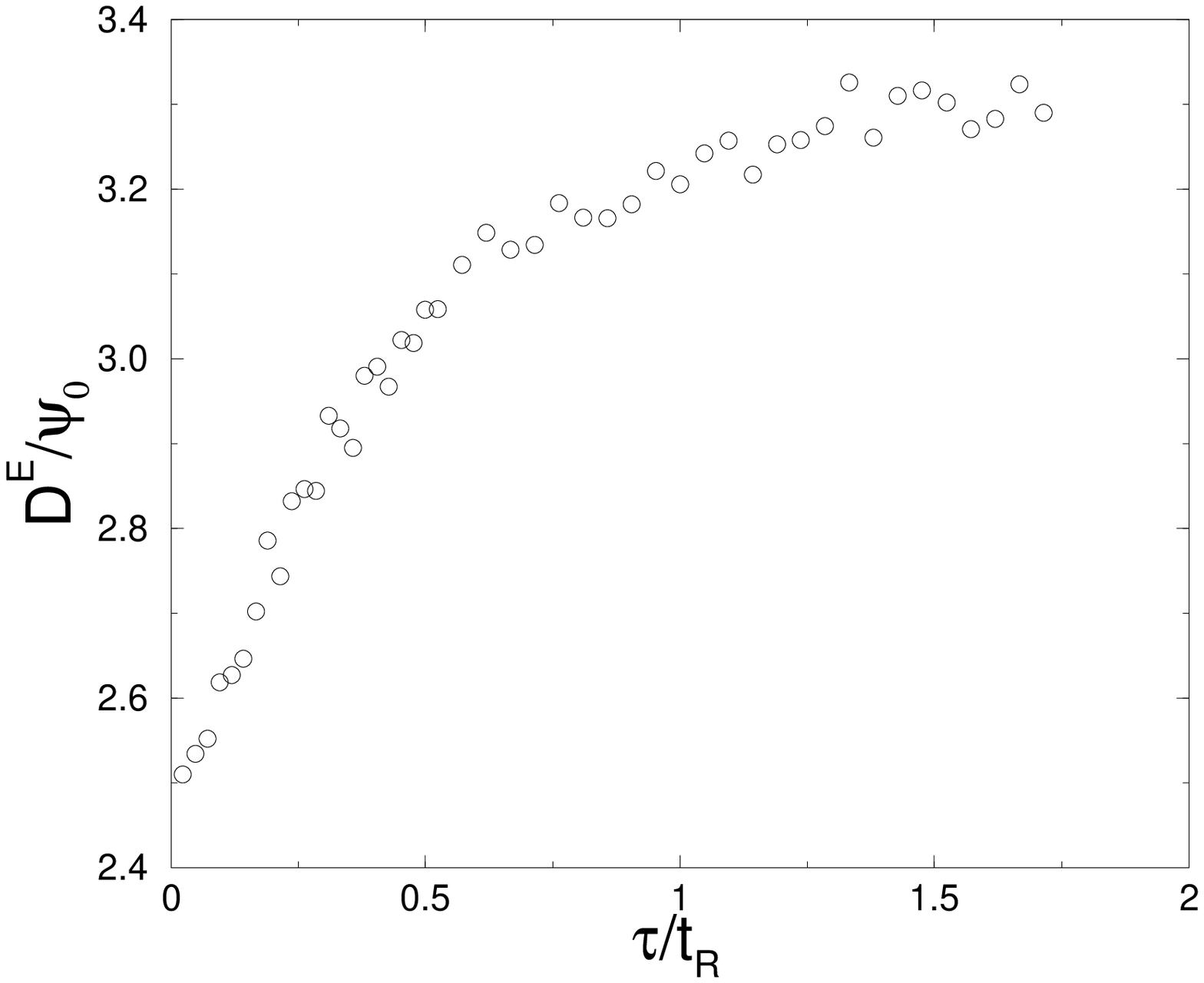,width=.9\linewidth}}
\end{center}
\end{minipage}
\caption{Behavior of the eddy-diffusivity
vs the correlation time of the small-scale velocity ${\bf v}'$.
On the left: $\omega/\omega_R$=0.67, corresponding to a local minimum
in the eddy-diffusivity profile of Fig.~1;
on the right: $\omega/\omega_R$=1.1, corresponding to the highest peak
(see again Fig.~1). In both cases,
$D_0/\psi_0=5\times 10^{-3}$.}
\end{figure}

The $x$-component, $D^E$, of the eddy-diffusivity {\it vs} the frequency,
$\omega$,
of the roll oscillation is shown in Fig.~1 for $D_0/\psi_0=5\times 10^{-3}$
and different values
of $\tau$: $\tau=0$ (full line), $\tau/t_R=0.24$ (dotted line),
$\tau/t_R=0.48$ (long-dashed line) and $\tau/t_R=0.95$ (dot-dashed line).
A few comments are in order.  Eddy-diffusivity
shows maxima originated from the resonance
between the lateral roll oscillation frequency and the characteristic
frequencies of the particle motion.
Moreover, the effect of $\tau$ on the shape of the peaks is twofold\,:
firstly, the variation of $\tau$ causes a shifting of maxima positions and,
secondly, the larger (smaller) $\tau$, the higher (lower) the peaks.
The first feature (particularly evident for $\omega$ corresponding to
the highest peak)
suggests that the correlation time, $\tau$, acts to renormalize
the large scale velocity. The result is
that an enhanced (with respect to the white-in-time case)
convective velocity governs the transport and, in order to have resonance
when increasing $\tau$, the roll oscillation frequency
must thus follow the increasing velocity.
The renormalizing effect of $\tau$
has been identified perturbatively in Ref.~\cite{CC98} for small $\tau$.
Our results
seem to suggest a  generalization for finite $\tau$.\\
Concerning the second effect played by $\tau$, this means that
peaks and valley of the eddy-diffusivity profile are associated
to regions of type WAR and EAR, respectively.
In terms of the two mechanisms associated to such regions,
the first is the winner for value
of $\omega$ corresponding to the peaks in the eddy-diffusivity profile
(where the contribution of anticorrelated regions is weak), while
the second dominates for values of $\omega$ corresponding to
local minima in the eddy-diffusivity profile (where  the weight of
anticorrelated regions is strong). This can be easily seen also
from Fig.~2 where behaviors of the eddy-diffusivity as a function of
$\tau$ are shown for two different value of $\omega$: $\omega/\omega_R=0.67$
(on the left) and $\omega/\omega_R=1.1$ (on the right). The former
value corresponds to a local minimum in the eddy-diffusivity profile,
while the latter to the highest peak (see Fig.~1).
\begin{figure}
\vfill \begin{minipage}{.495\linewidth}
\begin{center}
\mbox{\psfig{figure=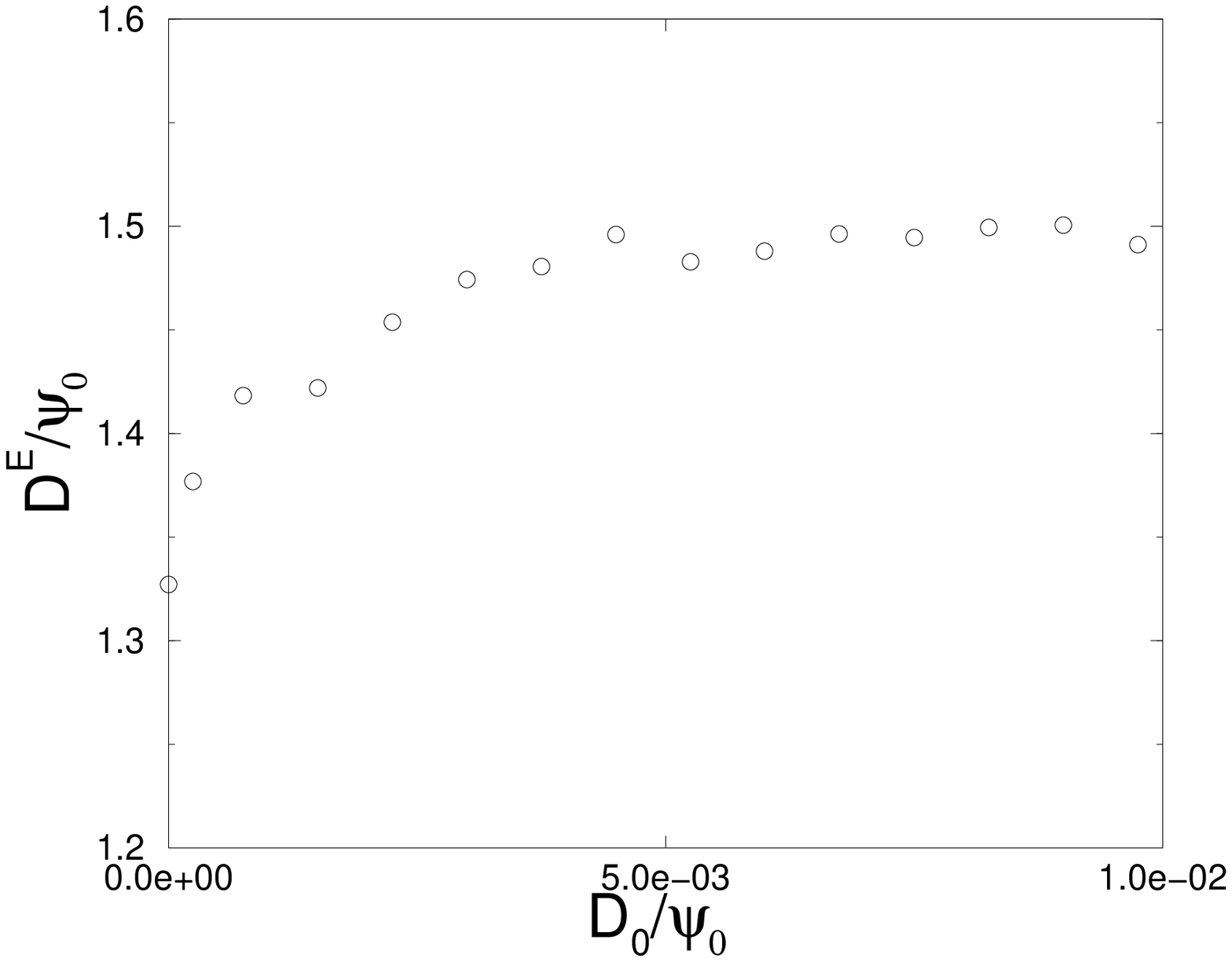,width=.9\linewidth}} 
\end{center}
\end{minipage} \hfill
\begin{minipage}{.495\linewidth}
\begin{center}
\mbox{\psfig{figure=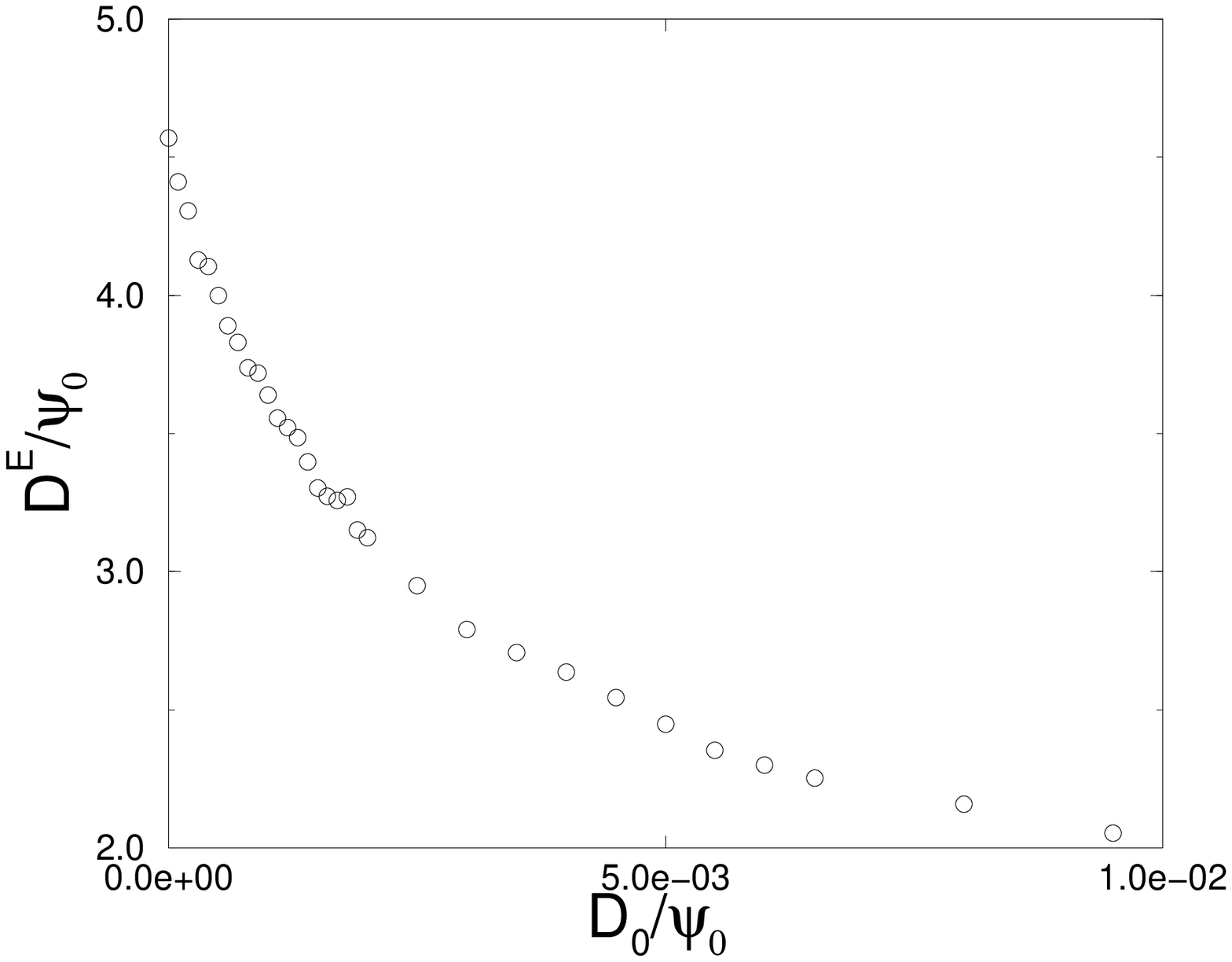,width=.9\linewidth}}
\end{center}
\end{minipage}
\caption{As in Fig.~2 but the eddy-diffusivity is plotted vs $D_0$
for a fixed value of the correlation time: $\tau=0$.}
\end{figure}

It is remarkable that the scenario above described is opposite when
fixing $\tau$ and varying the (small-scale) bare eddy-diffusivity
$D_0$. This can be easily observed from Fig.~3 where behaviors of the
eddy-diffusivity are now shown as a function of $D_0$, keeping $\tau$ fixed
and equal to zero. As in Fig.2, on the left we have $\omega/\omega_R=0.67$
while, on the right, $\omega/\omega_R=1.1$. The physical reason of such
behavior can be easily grasped from the aforesaid mechanisms but now
recalling the fact that an increasing $D_0$ makes the particles of
diffusing substance forget their past more rapidly (rather than less rapidly
as it happens by increasing $\tau$). This effect causes a reduction in the
transport. On the other hand, concerning the second mechanism,
trapping due to anticorrelated regions is now less effective. Indeed,
leaving the region of trapping is easier when increasing $D_0$. This fact
leads to
a reduction of the weight of the negative contribution to the time-integral
(\ref{Tayl})
giving the eddy-diffusivity and, as a consequence, transport is enhanced.\\
The final result is thus a complete symmetry between the following
operation: increasing $\tau$ (for a fixed $D_0$) and decreasing $D_0$
(for a fixed $\tau$). Large-scale
transport seems thus controlled by a parameter $\propto D_0/\tau$.
Roughly speaking, this is easily understood if we
observe that when increasing $\tau$ the particle motion along the Lagrangian
trajectories becomes more and more coherent. Conversingly, coherence
becomes lost when, for a fixed $\tau$, we decrease $D_0$. In this sense,
the dependence of Lagrangian trajectories on the statistical properties
of the small scale velocity is crucial in this problem.

In conclusion,  two ubiquitous different regimes of transport have been
identified here as relevant in the time-periodic roll circulation.
A key role to select such two regimes of transport enhancement/reduction
is due to the interplay between
the even oscillatory instability of the cell and the statistical properties
of the small-scale (random) velocity (e.g.~their correlation time or their
energy). Specifically,
in the model here considered, the even instability
is accounted for by a sinusoidal lateral boundary oscillation
with frequency $\omega$, while small-scale velocity activity is described
by a bare diffusivity, $D_0$, within
a Gaussian, zero-mean random process with correlation time $\tau$.
When varying $\omega$, the eddy diffusivity profile appears very
structured with sharp peaks separated by evident valley.
Peaks turn out to be associated to transport enhancement when
increasing $\tau$ (for a fixed $D_0$) or, conversingly,
when reducing $D_0$  (for a fixed $\tau$). The situation is opposite
for values of $\omega$ corresponding to minima in the eddy-diffusivity
profile. The physical key role is played by synchronization mechanisms
(from which the structured eddy-diffusivity profile arises) and by
the strength of the anticorrelated regions of the velocity field,
the weight of which in the time-integral giving the eddy-diffusivity
can be either enhanced (by reducing $D_0/\tau$) or reduced
(by increasing $D_0/\tau$), thus affecting in different ways the
large-scale transport.

\vskip 0.2cm
{\bf Acknowledgements}
We thank C.F.~Ratto, M.~Vergassola and A.~Vulpiani
for  several perceptive comments and suggestions on this work.
Useful suggestions and comments during
the 1999 TAO Study Center are also acknowledged (A.M.).
Simulations were performed at CINECA (INFM Parallel Computing Initiative).
P.C. has been supported by INFM (Progetto Ricerca
avanzata-Turbo) by MURST (program no. 9702265437) and
by the European grant ERB 4061 PL 97-0775.

\end{document}